\documentclass[sigconf]{acmart}

\usepackage{etoolbox}
\usepackage{booktabs}
\usepackage{algorithm}
\usepackage[noend]{algpseudocode}

\AtBeginDocument{%
  }

\newcommand{\paperTableStyle}{%
  \small
  \setlength{\tabcolsep}{4pt}%
  \renewcommand{\arraystretch}{1.08}%
}

\AtBeginEnvironment{table}{\paperTableStyle}
\AtBeginEnvironment{table*}{\paperTableStyle}
\AtBeginEnvironment{algorithm}{\paperTableStyle}



\begin{document}
\title[Beyond Metadata]{Beyond Metadata: CAPRA for Hidden Subgroup Analysis under Missing Metadata in Medical Imaging}

\author{
  Yawen Li,
  Yan Li,
  Zhe Xue,
  Yingxia Shao,
  Meiyu Liang,
  Guanhua Ye\textsuperscript{*}
}
\affiliation{%
  \institution{Beijing University of Posts and Telecommunications}
  \city{Beijing}
  \country{China}
}
\email{{warmly0716, liyanly, xuezhe, shaoyx, meiyu1210, g.ye}@bupt.edu.cn}
\thanks{*Corresponding author.}

\begin{abstract}
Medical imaging models are often deployed without the demographic, acquisition, and quality metadata needed for subgroup auditing. Once those metadata disappear, clinically critical failure modes can be masked by strong aggregate performance, and many robust-learning methods lose the group structure they rely on. We present CAPRA, a calibrated proxy-axis framework for hidden subgroup analysis under missing metadata. CAPRA predicts image-derived semantic axes, calibrates axis posteriors on a small metadata-labeled split via patient-level cross-fitting, and organizes those posteriors into a calibrated subgroup interface that supports both deployment-time failure analysis and downstream robust learning without requiring subgroup labels at deployment. Across fundus, dermoscopy, and chest radiography, CAPRA reveals disparity patterns missed by metadata-only slicing, remains informative under dataset shift, and produces subgroup partitions that align more closely with explicit failure axes than image-only or latent-slice baselines. The same interface can also be reused by downstream robust learners, although those gains are domain-dependent. Overall, CAPRA turns hidden subgroup analysis under missing metadata into a calibrated, interpretable, and reusable subgroup interface for deployment-time analysis and robust transfer.
\end{abstract}

\maketitle

\section{Introduction}
Medical imaging models are usually developed on curated research cohorts with relatively complete demographic, device, and acquisition metadata. Clinical deployment looks different. Patient demographics may be missing from imaging workflows, device provenance may be inconsistently recorded, and formal quality annotations may not exist at all. Public datasets often reflect the same documentation gap: metadata are incomplete, uneven across attributes, or simply not rich enough for deployment-facing subgroup audit \cite{paul2022demographic,galanty2024assessing,khan2021global,li2013gaussian,li2022federated}. The consequence is familiar but still under-addressed. A model can look reliable in aggregate while failing on clinically meaningful subsets defined by technical, anatomical, or phenotypic variation, and those failures may remain invisible precisely because the labels needed to audit them are unavailable \cite{oakden2020hidden,seah2021comprehensive}. Recent work on subgroup-shift monitoring further shows that clinically harmful subgroup prevalence changes can remain in-distribution at the image level and therefore escape conventional OOD detection, motivating population-level auditing at deployment \cite{koch2022hidden,koch2024distribution}. Classic studies in chest radiography and related clinical imaging tasks have shown that apparent model success can be driven by site-specific cues, healthcare-process variables, or subgroup dependent underdiagnosis \cite{zech2018variable,badgeley2019deep,seyyed2021underdiagnosis}. Related work further shows that demographic prediction and subgroup disparity can be mediated by confounders and acquisition parameters rather than directly by pathology alone \cite{duffy2022confounders,glocker2023algorithmic,lotter2024acquisition}.

Current approaches only partly close this gap. Group-robust methods help when the relevant groups are known; subgroup discovery methods can expose latent slices, but often without making them clinically auditable. Both assumptions are brittle in medical imaging, where available metadata are often incomplete or too coarse for deployment-facing audit, and purely latent groups may reveal heterogeneity without telling clinicians what failed or why \cite{paul2022demographic}. Beyond GroupDRO, JTT, and DFR, recent weak- or no-group methods such as failure-based debiasing, automatic feature reweighting, selective last-layer retraining, and group-aware priors further relax annotation requirements, but they still do not expose an interpretable subgroup interface for deployment-facing audit \cite{nam2020learning,qiu2023simple,labonte2023towards,rudner2024mind}. Work on incomplete multi-view multi-label learning also shows that sharing structure across partially observed views and label categories can reduce brittleness under missing observations \cite{ou2024view}, but that setting is oriented toward predictive completion rather than deployment-facing subgroup auditing. Medical vision-language work suggests that image variation can be organized into semantically meaningful concepts, but that semantic structure has rarely been turned into a calibrated subgroup interface under missing metadata \cite{huang2021gloria,wang2022medclip,zhang2023biomedclip}. This gap matters even more for medical vision-language models: recent studies show that semantic alignment can reveal hidden image-attribute structure, but it can also encode demographic information and amplify subgroup disparities \cite{kumar2025leveraging,luo2024fairclip,yang2025demographic}. What is missing is not another generic robust learner. It is an audit coordinate system that still works when the metadata needed to define groups have disappeared.

We therefore ask a paired question: can a model expose clinically meaningful subgroup failure when deployment metadata are missing, and can the resulting subgroup structure also support downstream robust learning? We answer that question with \textbf{CAPRA} (\textbf{C}alibrated \textbf{P}roxy-\textbf{A}xis \textbf{R}isk \textbf{A}uditing). CAPRA predicts image-derived proxy axes, calibrates their posterior probabilities on a small metadata-labeled cohort using patient-level cross-fitting, and uses those calibrated posteriors to construct a calibrated subgroup interface for failure analysis and robust learning without subgroup labels at deployment.

Our contributions are threefold. We recast hidden subgroup analysis under missing metadata as calibrated proxy-axis risk estimation rather than brittle recovery of sparse oracle groups. We introduce CAPRA, which combines proxy-axis prediction, posterior calibration, and trust-weighted axis selection to expose interpretable failure modes and provide a reusable subgroup interface. Finally, across fundus, dermoscopy, and chest X-ray, we show that CAPRA remains informative under external handheld fundus shift, aligns more closely with explicit failure axes than latent alternatives, and supports downstream robust learners in a domain-dependent way.

\section{Related Work}
\subsection{Hidden Stratification and Group-Robust Learning}
Hidden stratification explains why medical imaging systems can perform well on average while still failing on clinically meaningful subsets \cite{oakden2020hidden,yang2024limits}. In practice, those failures often concentrate on technical, anatomical, or demographic slices and are driven by shortcut cues such as site-specific artefacts, healthcare-process variables, or subgroup-dependent underdiagnosis \cite{zech2018variable,badgeley2019deep,seyyed2021underdiagnosis}. The challenge is that clinically relevant failure modes are not always aligned with the metadata most readily available to developers \cite{paul2022demographic,olesen2024slicing}.

Group-robust learning addresses this setting when subgroup structure is known or can be approximated with enough supervision. GroupDRO provides the canonical worst-group objective \cite{sagawa2019distributionally,li2008robust}, while JTT, DFR, and more recent reweighting methods reduce the label burden through harder-example mining, last-layer retraining, or bilevel weighting \cite{liu2021just,kirichenko2022last,qiao2025group}. CAPRA does not replace these methods. It addresses the missing interface they typically assume away: how to expose deployment-relevant subgroup structure when the metadata needed to define groups are incomplete or unavailable \cite{huo2023t2}.

\subsection{Subgroup Discovery without Metadata}
Recent work has also tried to discover subgroups directly from learned representations or model behavior, including GEORGE/No Subclass Left Behind, Domino, and later medical slice-discovery analyses \cite{sohoni2020no,eyuboglu2022domino,olesen2024slicing}. Recent medical work also frames subgroup discovery as a performance-monitoring problem, asking whether discovered slices expose larger clinically relevant disparities than conventional metadata-based subgroup analysis \cite{bissoto2025subgroup,koch2024distribution}. These methods are attractive because they suggest that systematic failure slices can sometimes be surfaced without explicit subgroup supervision. Related approaches such as ExMap and DPE move beyond raw embeddings by clustering explanation maps or diversifying prototype heads \cite{chakraborty2024exmap,to2025diverse}.

Their limitation is not that they fail to find heterogeneity, but that the resulting groups may still be clinically opaque or loosely connected to downstream robust optimization. CAPRA sits nearby, but with a different emphasis. It uses inferred subgroup structure too, yet keeps that structure tied to interpretable proxy axes so the output remains auditable rather than merely latent.

\begin{figure*}[t]
\centering
\includegraphics[width=0.8\textwidth]{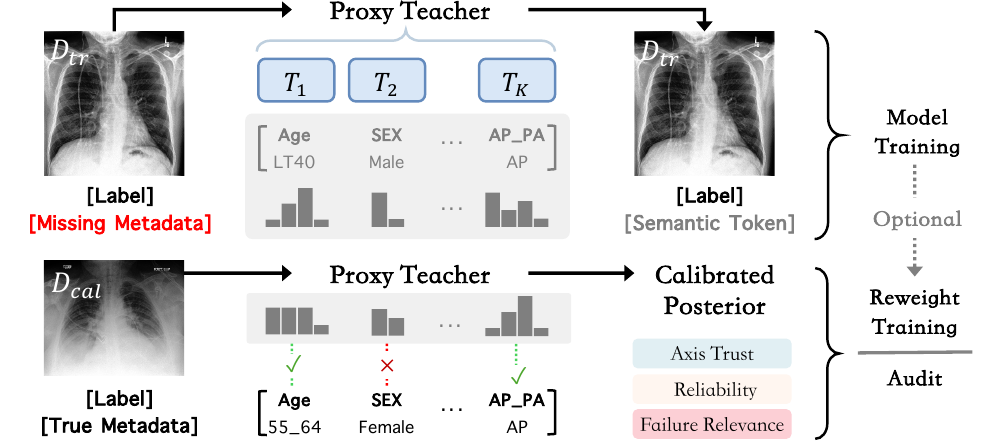}
\caption{Overview of CAPRA. Proxy teachers extract semantic axes from images to generate tokens for model training. A small metadata-labeled set calibrates proxy predictions into reliable posteriors, which are used to estimate axis importance and support metadata-free subgroup auditing and optional robust reweighting.}
\Description{A full-width overview figure for CAPRA. The left side shows the missing-metadata setting with training data, a small metadata-labeled calibration split, and unavailable deployment metadata. The middle shows proxy-axis teachers, hard semantic tokens, calibrated soft posteriors, and a warm-start predictor. The right side shows a calibrated subgroup interface reused for subgroup analysis and downstream robust learning.}
\label{fig:capra_overview}
\end{figure*}

\subsection{Representation Learning, Spurious Correlations, and Multimodal Semantics}
Another line of work studies why standard models fail on minority groups in the first place. Analyses of spurious correlations show that ERM can rely on shortcut features even when its representation still contains more stable task signal \cite{kirichenko2022last,qiao2025group}. That observation motivates post-hoc reweighting and probing, and it matters here for a simple reason: subgroup signal may already exist in the representation, but it still has to be exposed and organized when metadata are missing.

Medical vision-language models suggest one way to do that. By aligning images with semantically meaningful text, they make it plausible to express latent variation such as acquisition style, anatomical differences, or image quality in a structured semantic form rather than only in an opaque embedding space \cite{huang2021gloria,wang2022medclip,zhang2023biomedclip,bannur2023learning,xu2013image,linmei2019heterogeneous,xiao2022retromae}. Specialty-specific ophthalmic multimodal models further illustrate the promise of semantic supervision and transfer, but they are not designed to output calibrated subgroup posteriors or deployment-facing audit axes \cite{du2024ret,silva2025foundation,shi2025multimodal,tan2024fairer,zhang2024generalist}. Our use of proxy axes is also related to concept-based intermediate supervision \cite{koh2020concept}. The distinction is that we do not treat semantic structure as an end in itself or as a generic multimodal gain. We use it as a calibrated bridge from images to auditable subgroup hypotheses under missing metadata.

\section{Problem Setting}
We study supervised prediction when deployment metadata are unavailable. Each example contains an image \(X \in \mathcal X\) and task label \(Y \in \mathcal Y\). We distinguish two objects. The first is explicit audit metadata \(A=(A_1,\ldots,A_K)\), where each \(A_k \in \mathcal V_k\) is a discrete axis such as age bucket, sex, acquisition style, or image quality. These metadata are visible only on a small calibration and audit cohort. The second is a set of proxy semantic axes \(Z=(Z_1,\ldots,Z_K)\), where each \(Z_k \in \mathcal V_k\) shares the same vocabulary as \(A_k\) but is inferred from the image. CAPRA does not assume that \(A_k\) can be recovered exactly from \(X\); instead, it learns proxy axes \(Z_k\) that are semantically anchored to \(A_k\) on the metadata-labeled subset.

During development we observe a large task-labeled training set \(D_{\mathrm{tr}}\) whose metadata are hidden:
\begin{equation}
D_{\mathrm{tr}}=\{(x_i,y_i)\}_{i=1}^n
\end{equation}
We also observe a small disjoint metadata-labeled calibration and audit set \(D_{\mathrm{cal}}\):
\begin{equation}
D_{\mathrm{cal}}=\{(x_j,y_j,a_j)\}_{j=1}^m,\qquad m\ll n.
\end{equation}
Finally, we use a validation set \(D_{\mathrm{val}}\) for early stopping and model selection. At deployment, metadata are unavailable.

For a predictor \(h:\mathcal X\to\Delta^{|\mathcal Y|-1}\) and bounded loss \(\ell(h(x),y)\in[0,1]\), we define the proxy-axis bucket risk
\begin{equation}
R^{\mathrm{proxy}}_{k,v}(h)=\mathbb E[\ell(h(X),Y)\mid Z_k=v]
\end{equation}
and the support-filtered worst-axis risk
\begin{equation}
R^{\mathrm{proxy}}_{\mathrm{worst}}(h)=\max_{1\le k\le K}\ \max_{v:\pi^{\mathrm{proxy}}_{k,v}\ge \pi_{\min}} R^{\mathrm{proxy}}_{k,v}(h).
\end{equation}
where \(\pi^{\mathrm{proxy}}_{k,v}=\mathbb P(Z_k=v)\) and \(\pi_{\min}>0\) is a support threshold used to exclude negligibly small buckets. Our goal is to reduce \(R^{\mathrm{proxy}}_{\mathrm{worst}}(h)\) while preserving average task performance
\begin{equation}
R_{\mathrm{avg}}(h)=\mathbb E[\ell(h(X),Y)].
\end{equation}

Direct optimization over the full Cartesian product \(Z_1\times\cdots\times Z_K\) is statistically unstable in the low-metadata regime because support collapses quickly as the number of axes grows. We therefore do not try to reconstruct a full oracle subgroup identity. CAPRA treats missing-metadata robustness as an axis-selective proxy-risk problem. Explicit metadata \(A\) are used to calibrate, align, and audit the learned proxy substrate; the primary training object is proxy-axis risk. This axis-selective formulation is compatible with multi-group views of hidden stratification that reason over conditional risks across many overlapping subgroups \cite{tosh2022simple}.

\section{Method}
CAPRA turns missing-metadata robustness into a reusable calibrated proxy-axis interface. Figure~\ref{fig:capra_overview} summarizes the full pipeline. It first constructs semantically anchored proxy axes, then calibrates their posterior probabilities under patient-level cross-fitting, and finally uses the resulting interface for failure-aware adaptation or downstream robust transfer. The main paper focuses on the interface objects that drive this pipeline and the mechanisms that make them useful under missing metadata.

\subsection{Calibrated Proxy-Axis Interface}
For each explicit audit axis \(k\), CAPRA trains or obtains a proxy teacher that produces logits \(s^{(k)}(x)\) over a shared vocabulary \(\mathcal V_k\). The hard proxy predictions are rendered into a compact structured token string
\begin{equation}
\begin{aligned}
t(x)&=\mathrm{Tok}\big(\hat a_1(x),\ldots,\hat a_K(x)\big),\\
\hat a_k(x)&=\arg\max_{v\in\mathcal V_k} s_v^{(k)}(x).
\end{aligned}
\end{equation}
Tokens such as \texttt{[AGE=55\_64]} and \texttt{[QUALITY=FAIR]} make subgroup structure explicit for the warm-start predictor \(h_0\), which is trained on \(D_{\mathrm{tr}}\) with a gated semantic-visual fusion module. CAPRA does not use the hard tokens alone for auditing or weighting; those stages operate on calibrated soft proxy posteriors.

The central object is therefore the calibrated proxy-axis posterior \(\hat q_{i,v}^{(k)} \approx P(Z_k=v\mid X_i)\). Following standard post-hoc calibration practice \cite{guo2017calibration}, CAPRA combines temperature scaling with confusion-matrix shrinkage under patient-level cross-fitting on the metadata-labeled calibration cohort:
\begin{equation}
\begin{aligned}
q_{i,v}^{(k)}
=&(1-\lambda_k)\,\mathrm{softmax}(s_i^{(k)}/\tau_k)_v\\
&+\lambda_k\, C_{k,\hat v_i,v},
\end{aligned}
\end{equation}
Here \(\hat v_i\) denotes the most likely bucket under the temperature-scaled teacher. We further define the sample-level reliability score
\begin{equation}
r_i^{(k)} = 1-\frac{H(q_i^{(k)})}{\log |\mathcal V_k|},
\end{equation}
with \(H(\cdot)\) denoting Shannon entropy. The calibrated interface is then \(\mathcal I=\{t,q,r,\alpha\}\): tokens provide semantic anchoring, posteriors preserve subgroup uncertainty, reliability suppresses unstable proxy assignments, and axis weights determine where robustness pressure is applied.

\begin{algorithm}[t]
\caption{CAPRA pipeline.}
\label{alg:capra_pipeline}
\begin{algorithmic}[1]
\Require \(D_{\mathrm{tr}}, D_{\mathrm{cal}}, D_{\mathrm{val}}, \{T_k\}_{k=1}^K, \pi_{\min}\)
\For{each sample \(x_i \in D_{\mathrm{tr}} \cup D_{\mathrm{cal}}\)}
    \For{each axis \(k=1,\ldots,K\)}
        \State infer logits \(s_i^{(k)} \gets T_k(x_i)\) and hard proxy bucket \(\hat a_k(x_i)\)
    \EndFor
    \State form structured token string \(t(x_i)=\mathrm{Tok}(\hat a_1(x_i),\ldots,\hat a_K(x_i))\)
\EndFor
\State train warm-start predictor \(h_0\) on \(D_{\mathrm{tr}}\) using CAPRA tokens
\For{each axis \(k=1,\ldots,K\)}
    \State fit \(\tau_k\) and \(C_k\) on \(D_{\mathrm{cal}}\) with patient-level cross-fitting
    \State compute calibrated posteriors \(q_i^{(k)}\), reliabilities \(r_i^{(k)}\), and trust \(u_k\)
    \State estimate support-filtered disparity \(\Delta_k(h_0)\) on \(D_{\mathrm{cal}}\)
\EndFor
\State normalize axis weights \(\alpha_k \propto u_k \Delta_k(h_0)\)
\State build calibrated interface \(\mathcal{I}=\{t, q, r, \alpha\}\)
\If{standalone CAPRA adaptation}
    \State initialize \(h \gets h_0\); freeze image and token encoders
    \For{epoch \(=1,\ldots,T\)}
        \For{mini-batch \(B_t \subset D_{\mathrm{tr}}\)}
            \State update EMA proxy-bucket risks \(\tilde R_{k,v}^{(t)}\)
            \State compute excess-risk scores \(e_i^{(k)}\) and normalized weights \(w_i\)
            \State update fusion and classifier layers using \(L_{\mathrm{adapt}}\)
        \EndFor
        \State keep checkpoints within one standard error of the best BA on \(D_{\mathrm{val}}\)
    \EndFor
    \State select the retained checkpoint with the best worst-axis BA on \(D_{\mathrm{cal}}\)
    \State \textbf{return} \(\mathcal{I}, h\)
\Else
    \State train the chosen downstream robust learner using \(\mathcal{I}\)
    \State \textbf{return} \(\mathcal{I}\) and the downstream model
\EndIf
\end{algorithmic}
\end{algorithm}

\subsection{Axis Weighting and Failure-Aware Adaptation}
Not every proxy axis should receive the same robustness mass. CAPRA combines an axis-trust score \(u_k\), derived from cross-fitted proxy quality, with a failure-relevance score \(\Delta_k(h_0)\), derived from the warm-start model's support-filtered proxy-axis disparity. Their normalized product defines the axis weight
\begin{equation}
\alpha_k
=
\frac{u_k\,\Delta_k(h_0)}
{\sum_{b=1}^K u_b\,\Delta_b(h_0)+\varepsilon}.
\end{equation}
This weighting concentrates pressure on axes that are both well calibrated and empirically tied to failure.

Standalone CAPRA then performs a short second stage initialized from \(h_0\), freezing the image and token encoders and updating only the fusion and classifier layers. For sample \(i\), CAPRA forms an excess-risk score \(e_i^{(k)}\) from calibrated proxy-bucket risk estimates on trustworthy axes and defines
\begin{equation}
\begin{aligned}
\hat w_i
&=
1+\lambda\sum_{k=1}^K \alpha_k r_i^{(k)} e_i^{(k)},\\
w_i
&=
\frac{\min(\rho,\hat w_i)}
{|B_t|^{-1}\sum_{j\in B_t}\min(\rho,\hat w_j)}.
\end{aligned}
\end{equation}
The second-stage objective is
\begin{equation}
L_{\mathrm{adapt}}
=
\frac{1}{|B_t|}
\sum_{i\in B_t} w_i\,\ell(h(x_i),y_i).
\end{equation}
In the transfer experiments, the same calibrated interface is passed to the downstream robust learner instead of running CAPRA's own second stage.

\subsection{Model Selection and Leakage Control}
Checkpoint selection follows a one-standard-error rule rather than a weighted composite score. We retain checkpoints whose validation balanced accuracy lies within one standard error of the best validation model, then select the one with the highest support-filtered worst-axis balanced accuracy on the calibration cohort, breaking ties by macro-F1. The rule keeps promotion transparent and reduces overfitting to a small metadata-labeled split. External audit data are never used for calibration, selection, or tuning.

Algorithm~\ref{alg:capra_pipeline} summarizes the full pipeline, with interface construction as the common backbone for both standalone adaptation and downstream transfer.

\section{Experiments}
\subsection{Research Questions}
We organize the experiments around six complementary questions covering downstream transfer, external generalization, cross-domain failure visibility, semantic alignment, calibration efficiency, and interface quality.

\textbf{RQ1. Consistency of downstream gains under missing metadata.} How consistently does the calibrated subgroup interface improve downstream robust learners under missing metadata?

\textbf{RQ2. External generalization under deployment shift.} Does the same calibrated interface remain informative under external shift?

\textbf{RQ3. Explicit failure-axis portability.} Which explicit failure axes remain visible across domains under standalone CAPRA?

\textbf{RQ4. Semantic alignment of discovered partitions.} Are CAPRA-induced subgroup partitions more semantically aligned with explicit failure structure than latent alternatives?

\textbf{RQ5. Calibration efficiency and axis weighting.} How much metadata-labeled calibration is needed, and which proxy axes receive robustness mass?

\textbf{RQ6. Interface quality under missing metadata.} Does the calibrated CAPRA subgroup interface add value on its own terms under missing metadata?

\begin{table*}[t]
\centering
\caption{Three-dataset comparison under missing metadata. \(\ast\) GroupDRO is included in its proxy-group variant and is less naturally matched to the missing-metadata setting.}
\label{tab:brset_main}
\small
\setlength{\tabcolsep}{3.5pt}
\renewcommand{\arraystretch}{1.08}
\begin{tabular*}{\textwidth}{@{\extracolsep{\fill}}lcccccccccccc}
\toprule
Method & \multicolumn{4}{c}{BRSET} & \multicolumn{4}{c}{HAM10000} & \multicolumn{4}{c}{CheXpert} \\
\cmidrule(lr){2-5}\cmidrule(lr){6-9}\cmidrule(lr){10-13}
 & BA (\%) & $\Delta$ & WGA (\%) & $\Delta$ & BA (\%) & $\Delta$ & WGA (\%) & $\Delta$ & BA (\%) & $\Delta$ & WGA (\%) & $\Delta$ \\
\midrule
\textbf{GroupDRO$^\ast$} & 62.6$\pm$2.1 & -- & 51.5$\pm$3.6 & -- & 33.8$\pm$2.7 & -- & 20.5$\pm$4.2 & -- & 65.6$\pm$1.3 & -- & 48.1$\pm$2.4 & -- \\
\quad + CAPRA & 83.1$\pm$1.4 & +20.5 & 69.9$\pm$2.0 & +18.4 & 72.9$\pm$1.7 & +39.1 & 33.2$\pm$3.3 & +12.7 & 82.5$\pm$0.8 & +16.9 & 61.4$\pm$1.9 & +13.3 \\
\addlinespace[2pt]
\textbf{JTT} & 73.2$\pm$2.6 & -- & 60.3$\pm$3.2 & -- & 56.1$\pm$2.0 & -- & 27.6$\pm$4.0 & -- & 81.0$\pm$1.2 & -- & 61.5$\pm$2.1 & -- \\
\quad + CAPRA & 77.9$\pm$1.5 & +4.7 & 73.2$\pm$2.3 & +12.9 & 69.5$\pm$2.1 & +13.4 & 41.0$\pm$2.7 & +13.4 & 78.6$\pm$0.9 & -2.4 & 66.9$\pm$1.7 & +5.4 \\
\addlinespace[2pt]
\textbf{DFR} & 72.9$\pm$1.6 & -- & 65.7$\pm$2.0 & -- & 68.7$\pm$1.4 & -- & 26.8$\pm$2.9 & -- & \textbf{89.5$\pm$0.7} & -- & \textbf{77.7$\pm$1.3} & -- \\
\quad + CAPRA & 74.3$\pm$1.1 & +1.4 & 70.4$\pm$1.9 & +4.7 & 69.0$\pm$1.5 & +0.3 & 28.0$\pm$2.1 & +1.2 & 81.0$\pm$0.9 & -8.5 & 74.5$\pm$1.2 & -3.2 \\
\addlinespace[2pt]
\textbf{DPE} & 84.6$\pm$1.2 & -- & 66.9$\pm$2.2 & -- & 71.9$\pm$1.6 & -- & 39.1$\pm$3.0 & -- & 84.6$\pm$1.0 & -- & 69.6$\pm$1.7 & -- \\
\quad + CAPRA & \textbf{86.0$\pm$0.7} & +1.4 & \textbf{79.0$\pm$1.7} & +12.1 & \textbf{74.5$\pm$1.1} & +2.6 & \textbf{49.4$\pm$2.4} & +10.3 & 85.8$\pm$0.6 & +1.2 & 71.5$\pm$1.8 & +1.9 \\
\addlinespace[2pt]
\textbf{GSR} & 76.8$\pm$1.9 & -- & 61.0$\pm$2.7 & -- & 58.6$\pm$2.3 & -- & 34.1$\pm$3.5 & -- & 78.7$\pm$1.1 & -- & 66.2$\pm$2.2 & -- \\
\quad + CAPRA & 72.5$\pm$1.4 & -4.3 & 64.4$\pm$2.6 & +3.4 & 58.3$\pm$1.7 & -0.3 & 41.1$\pm$2.9 & +7.0 & 80.3$\pm$1.2 & +1.6 & 68.3$\pm$1.6 & +2.1 \\
\bottomrule
\end{tabular*}
\end{table*}

\subsection{Datasets}
We evaluate CAPRA on four public datasets chosen to separate internal fundus evidence, external handheld audit, and cross-domain transfer beyond fundus imaging.

\textbf{BRSET} is the headline internal fundus dataset \cite{nakayama2023brazilian}. We use it for three-class diabetic retinopathy grading, keep referable diabetic retinopathy as auxiliary evidence rather than a headline task, audit age bucket, sex, camera, focus, illumination, image field, and artifacts, and report balanced accuracy plus support-filtered worst-group accuracy on patient-level splits.

\textbf{mBRSET} is used only as the external handheld fundus audit target \cite{nakayama2024mbrset}. Its labels are harmonized to the same three-class diabetic retinopathy audit contract as BRSET, and we report external balanced accuracy and Macro-F1 without using mBRSET for calibration, tuning, or checkpoint selection.

\textbf{HAM10000} is the cross-domain dermoscopy validation dataset \cite{tschandl2018ham10000}. The headline task is seven-class lesion classification, the explicit audit axes are age bucket, sex, and lesion localization, and the main reported metrics are balanced accuracy and support-filtered worst-group accuracy on the fixed lesion-level split.

\textbf{CheXpert} is the cross-domain chest X-ray validation dataset \cite{irvin2019chexpert}. We use Pleural Effusion as a binary target, audit sex, age bucket, frontal/lateral view, and AP/PA protocol, and report balanced accuracy and support-filtered worst-group accuracy on the official patient-level split.

Across all datasets, true metadata are hidden during task training and reserved for calibration, alignment, and subgroup audit. Even the GroupDRO comparator is implemented in a proxy-group variant rather than an oracle-label setting. This distinction is essential because the paper studies deployment-facing subgroup auditing when the metadata needed to define groups are unavailable at training or target time.

\begin{figure*}[t]
\centering
\includegraphics[width=\textwidth]{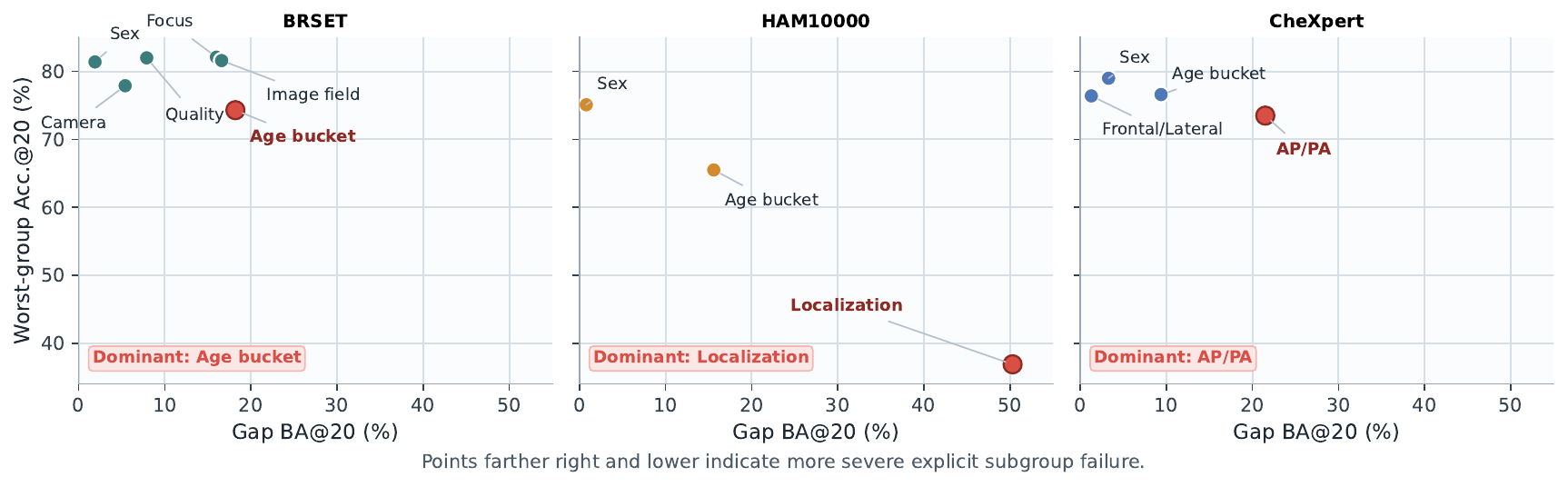}
\caption{Explicit subgroup failure map for standalone CAPRA (CAPRA-S) across BRSET, HAM10000, and CheXpert. Each point is an explicit slice axis positioned by its support-filtered Gap BA@20 and the corresponding support-filtered Worst-group Acc.@20; the red point marks the dominant failure axis in each domain.}
\Description{A three-panel explicit subgroup failure map compares BRSET, HAM10000, and CheXpert for standalone CAPRA. In each panel, each slice axis is shown as a labeled point using Gap BA at support threshold 20 on the horizontal axis and Worst-group Accuracy at support threshold 20 on the vertical axis. The dominant failure axis in each domain is highlighted in red: age bucket for BRSET, localization for HAM10000, and AP/PA for CheXpert. Points farther right and lower indicate more severe explicit subgroup failure.}
\label{fig:explicit_failure_map}
\end{figure*}

Throughout the paper, \textbf{WGA} denotes support-filtered worst-group accuracy under each task's default support rule. By contrast, \textbf{Worst-group Acc.@20} in Figure~\ref{fig:explicit_failure_map} fixes the support threshold at 20 and is used only for explicit failure localization.

\subsection{Baselines}
We compare our method against six representative approaches that span the major paradigms in subgroup robustness, enabling interpretable attribution of performance gains.

\textbf{GroupDRO} is included here in a proxy-group, missing-metadata variant that applies the same worst-group objective over discovered proxy groups rather than oracle subgroup labels \cite{sagawa2019distributionally}.

\textbf{JTT} (Just Train Twice) is a two-stage reweighting method that treats samples misclassified by early ERM training as proxies for minority groups \cite{liu2021just}.

\textbf{DFR} (Deep Feature Reweighting) assumes that standard ERM already learns useful core representations and improves robustness by retraining only the final classifier on balanced data \cite{kirichenko2022last}.

\textbf{ExMap} belongs to the subgroup discovery paradigm and infers pseudo-groups by clustering explainability heatmaps instead of raw feature representations \cite{chakraborty2024exmap}.

\textbf{DPE} (Diverse Prototypical Ensembles) replaces a single classifier with an ensemble of diversified prototype classifiers to capture multiple latent decision boundaries without requiring predefined subgroup labels \cite{to2025diverse}.

\textbf{GSR} (Group-robust Sample Reweighting) combines last-layer retraining with influence-function-based sample reweighting to improve robustness under limited subgroup supervision \cite{qiao2025group}.

These baselines are selected because they collectively cover explicit group-based robustness, weakly supervised reweighting, post-hoc feature reweighting, and unsupervised subgroup discovery.

\begin{table*}[t]
\centering
\caption{Matched BRSET and external mBRSET evaluation of CAPRA.}
\label{tab:brset_audit}
\begin{tabular*}{\textwidth}{@{\extracolsep{\fill}}lcccc}
\toprule
Method & BRSET audit BA (\%) & BRSET audit WGA (\%) & mBRSET BA (\%) & mBRSET Macro-F1 (\%) \\
\midrule
CAPRA anchor & \textbf{81.6 $\pm$ 1.6} & \textbf{72.9 $\pm$ 2.2} & 66.8 $\pm$ 3.1 & 52.6 $\pm$ 3.7 \\
CAPRA (ours) & 80.2 $\pm$ 2.0 & 71.1 $\pm$ 2.5 & \textbf{69.1 $\pm$ 2.7} & \textbf{60.8 $\pm$ 4.2} \\
\bottomrule
\end{tabular*}
\end{table*}

\section{Results}
\subsection{RQ1. Consistency of Downstream Gains under Missing Metadata}
Table~\ref{tab:brset_main} shows broad downstream gains under missing metadata: CAPRA improves WGA in 14 of 15 learner-dataset comparisons and BA in 11 of 15.

The pattern is strongest in WGA rather than BA. On BRSET and HAM10000, the best overall results come from DPE + CAPRA at 79.0 $\pm$ 1.7 and 49.4 $\pm$ 2.4 WGA, while CheXpert marks the boundary, where DFR remains strongest at 89.5 $\pm$ 0.7 BA and 77.7 $\pm$ 1.3 WGA. This WGA-first behavior is the first signal of what CAPRA is adding: the calibrated interface does not primarily sharpen average-case discrimination, but rather gives downstream learners a more stable handle on the minority slices where their errors concentrate. DPE still benefits even though it already models latent decision diversity, which suggests that CAPRA is not merely rescuing weak baselines; the proxy-axis interface contributes semantically aligned subgroup structure that latent diversity alone does not fully recover. CheXpert then clarifies the limit: when a strong learner such as DFR already captures much of the dominant subgroup structure, the remaining room for CAPRA to improve is smaller, and average-case metrics can become less responsive than worst-group ones. Taken together, RQ1 suggests that CAPRA is most useful when the downstream learner still lacks a stable, semantically anchored view of hidden subgroup structure.

\subsection{RQ2. External Generalization under Deployment Shift}
Table~\ref{tab:brset_audit} shows a split pattern between matched BRSET and external handheld mBRSET.

On matched BRSET, the simpler CAPRA anchor remains slightly stronger at 81.6 $\pm$ 1.6 BA and 72.9 $\pm$ 2.2 WGA. Under external handheld shift, that ordering reverses: the full model improves mBRSET from 66.8 $\pm$ 3.1 to 69.1 $\pm$ 2.7 BA and from 52.6 $\pm$ 3.7 to 60.8 $\pm$ 4.2 Macro-F1. This reversal suggests that calibration and failure-aware weighting are not merely tightening the fit to the source audit task; they preserve subgroup signal better once acquisition conditions change. The more structured variant is therefore not always the best in-domain scorer, but it is the more transferable subgroup estimator, which is precisely the deployment regime that matters under missing metadata.

Under shift, the extra weighting layer becomes useful because acquisition changes alter which slices are genuinely hard. In that regime, calibration and weighting act less like source-performance enhancers and more like a buffer against source specific subgroup shortcuts.

\subsection{RQ3. Explicit Failure-Axis Portability}
Figure~\ref{fig:explicit_failure_map} compares explicit failure axes for standalone CAPRA-S, where CAPRA-S denotes the standalone CAPRA model (`CAPRA (ours)' in Table~\ref{tab:brset_audit}).

Two anchors make the portability boundary clear: age bucket leads BRSET with an 18.2\% Gap BA@20, while localization dominates HAM10000 at 50.3\%. CAPRA is therefore not recovering a universal subgroup taxonomy; it is exposing whichever proxy axis is most tightly coupled to failure under the dominant acquisition and phenotype variation of each modality. That is the portability claim supported here: the interface transfers, but the dominant failure semantics do not have to. In practice, cross-domain evaluation should therefore ask whether failure remains visible, not whether the same named axis repeats.

Age on BRSET points to a phenotype-linked failure mode, whereas localization on HAM10000 points to a spatial or lesion-context failure mode. A scalar WGA can show that hidden subgroup risk exists, but it cannot show whether that risk is driven by age, acquisition, anatomy, or lesion context. The portability evidenced here is thus portability of the auditing interface, not portability of the dominant slice itself.

\subsection{RQ4. Semantic Alignment of Discovered Partitions}
\begin{figure*}[t]
\centering
\includegraphics[width=\textwidth]{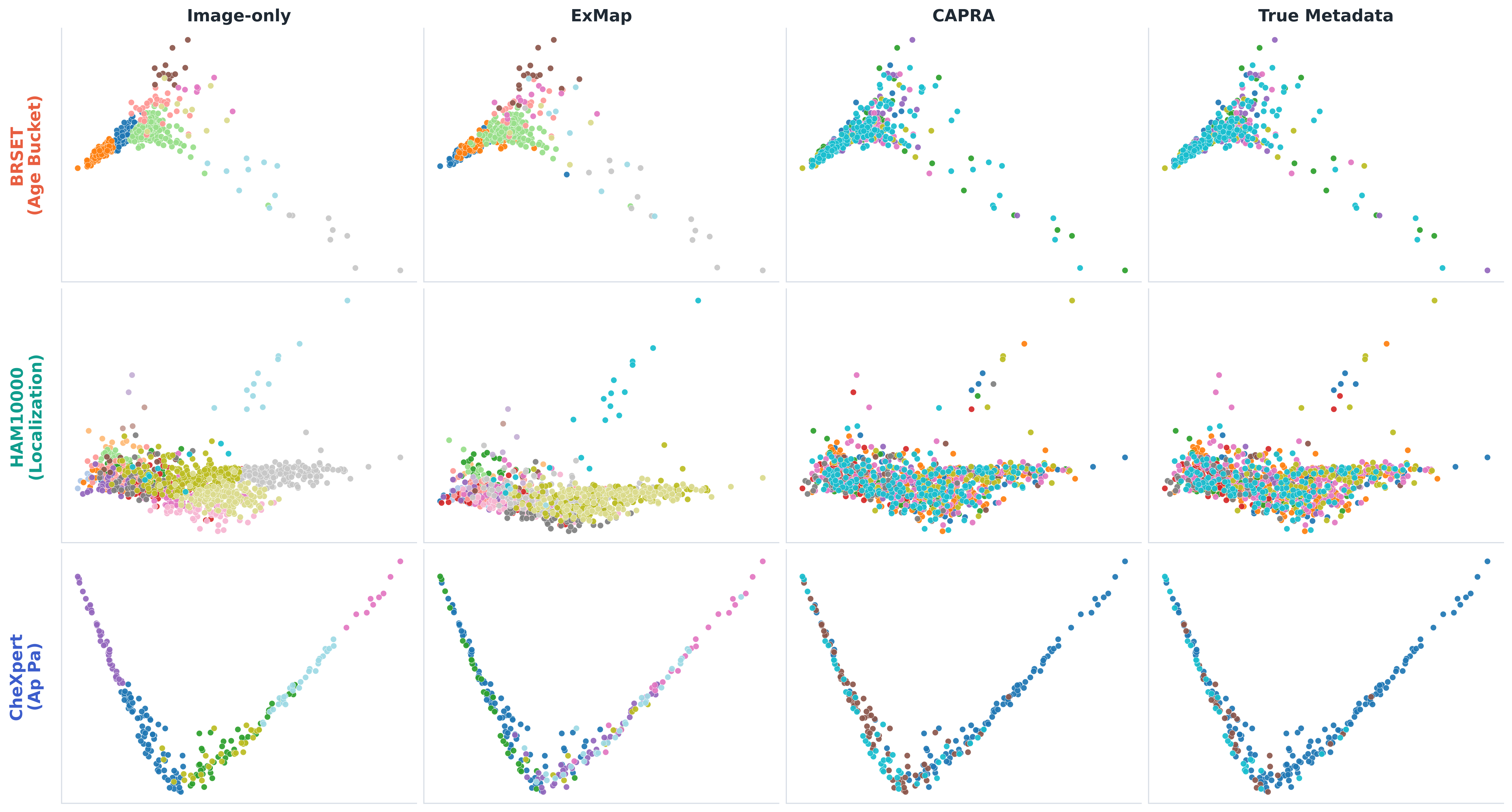}
\caption{Aligned 2D subgroup partitions across BRSET, HAM10000, and CheXpert. Within each dataset, Image-only, ExMap, CAPRA, and True Metadata partitions are shown on the same embedding to visualize semantic alignment with the dominant explicit failure axis.}
\Description{A 3 by 4 grid of 2D subgroup visualizations. Rows correspond to BRSET, HAM10000, and CheXpert. Columns correspond to Image-only, ExMap, CAPRA, and True Metadata partitions. Within each row the same 2D coordinates are reused across all four columns so the partitions are visually comparable.}
\label{fig:rq4_projection_companion}
\end{figure*}

Figure~\ref{fig:rq4_projection_companion} and Table~\ref{tab:rq4_projection_companion} show that CAPRA aligns more closely with explicit failure structure than image-only or ExMap partitions.

\begin{table*}[t]
\centering
\caption{Quantitative companion to Figure~\ref{fig:rq4_projection_companion}. Pur./NMI denotes majority-mapped purity / normalized mutual information, Agr. denotes agreement with the dominant explicit failure axis, and Gap denotes the fixed-anchor subgroup gap. Agr. and Gap are reported in percent.}
\label{tab:rq4_projection_companion}
\setlength{\tabcolsep}{2pt}
\begin{tabular*}{\textwidth}{@{\extracolsep{\fill}}lccccccccc}
\toprule
Source & \multicolumn{3}{c}{BRSET} & \multicolumn{3}{c}{HAM10000} & \multicolumn{3}{c}{CheXpert} \\
\cmidrule(lr){2-4}\cmidrule(lr){5-7}\cmidrule(lr){8-10}
 & Pur./NMI $\uparrow$ & Agr. (\%) $\uparrow$ & Gap (\%) $\downarrow$ & Pur./NMI $\uparrow$ & Agr. (\%) $\uparrow$ & Gap (\%) $\downarrow$ & Pur./NMI $\uparrow$ & Agr. (\%) $\uparrow$ & Gap (\%) $\downarrow$ \\
\midrule
\textbf{True metadata} & \textbf{0.98} / \textbf{1.00} & \textbf{100.0} & \textbf{4.9} & \textbf{0.95} / \textbf{1.00} & \textbf{100.0} & \textbf{33.4} & \textbf{0.98 }/ \textbf{1.00} & \textbf{100.0} & \textbf{6.3} \\
\midrule
Image-only & 0.24 / 0.08 & 37.6 & 38.6 & 0.21 / 0.12 & 29.1 & 51.0 & 0.31 / 0.11 & 72.2 & 81.3 \\
ExMap & 0.08 / 0.06 & 37.4 & 36.3 & 0.23 / 0.11 & 27.8 & 51.4 & \textbf{0.61} / 0.33 & \textbf{84.2} & 63.1 \\
CAPRA & \textbf{0.81} / \textbf{0.60} & \textbf{83.3} & \textbf{19.0} & \textbf{0.68} / \textbf{0.68} & \textbf{81.7} & \textbf{10.8} & 0.55 / \textbf{0.43} & 70.5 & \textbf{22.5} \\
\bottomrule
\end{tabular*}
\end{table*}

The main positive anchor is BRSET, where CAPRA reaches 0.81 / 0.60 Pur./NMI, 83.3\% agreement, and a 19.0 subgroup gap. CheXpert provides the counterpoint: ExMap attains higher AP/PA agreement, 84.2\% versus 70.5\%, but its subgroup gap is far larger, 63.1 versus 22.5. Agreement alone is therefore insufficient. The proxy-axis design matters because it preserves semantic alignment without washing out the disparity signal that makes subgroup analysis useful.

CAPRA's semantic anchoring constrains the partitioning process by encouraging interpretability while still retaining clusters according to error concentration rather than visual coherence alone. The CheXpert counterexample makes this distinction sharp: ExMap can recover a partition that looks more faithful to one explicit acquisition label under majority mapping, yet still spread the true error mass too diffusely to be operationally useful. CAPRA's softer axis-based representation avoids that failure mode by preserving partial subgroup structure instead of forcing every sample into a single latent bucket.

\subsection{RQ5. Calibration Efficiency and Axis Weighting}
\begin{figure}[t]
\centering
\includegraphics[width=\columnwidth]{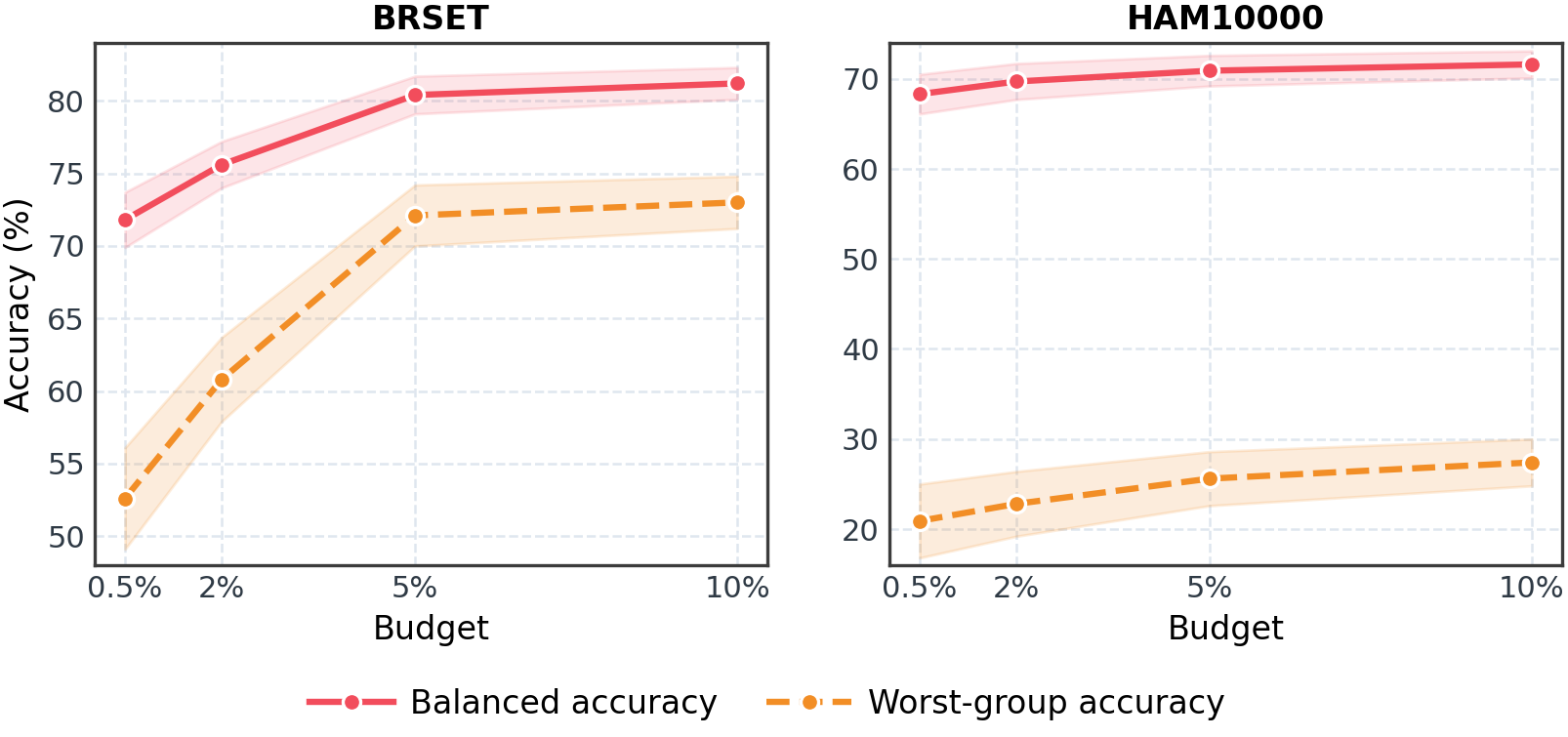}
\caption{Calibration-budget sweep for standalone CAPRA on BRSET and HAM10000.}
\Description{A single-column figure with two side-by-side line charts shows CAPRA calibration-budget trends for BRSET and HAM10000. In each panel, a solid red balanced-accuracy curve and a dashed orange worst-group-accuracy curve rise as calibration budget increases from 0.5 percent to 10 percent, with translucent bands indicating one standard deviation around each curve. BRSET shows a larger gain than HAM10000.}

\label{fig:budget_sweep_completed}
\end{figure}

Figure~\ref{fig:budget_sweep_completed} shows how quickly calibration improves with additional metadata, while the learned axis-diagnostic pattern indicates which axes receive robustness mass.

The main budget anchor is BRSET, where most gains arrive between 2\% and 5\% before the curves level off. The second anchor comes from the learned weighting pattern: QUALITY carries the highest trust, while AGE carries the highest failure relevance and normalized weight. A modest amount of metadata first stabilizes the calibrated subgroup signal, after which CAPRA routes robustness mass toward axes that are both reliable and failure-relevant.

\subsection{RQ6. Interface Quality under Missing Metadata}
Table~\ref{tab:ham_substrate_transfer} shows consistent gains from the CAPRA interface across all four learner-dataset blocks.

\begin{table}[t]
\centering
\caption{Reusable-interface ablation for JTT and DPE under matched subgroup audit.}
\label{tab:ham_substrate_transfer}
\small
\setlength{\tabcolsep}{3.5pt}
\begin{tabular}{llcc}
\toprule
Learner & Setting & BA (\%) & WGA (\%) \\
\midrule
\multicolumn{4}{l}{\textbf{BRSET}} \\
\textbf{JTT} & Image-only anchor & 63.7$\pm$2.3 & 58.9$\pm$3.8 \\
& CAPRA interface & 73.9$\pm$1.6 & 70.8$\pm$2.4 \\
& \shortstack[l]{CAPRA +\\ calibrated priors} & \textbf{75.2$\pm$1.4} & \textbf{73.6$\pm$2.0} \\
\addlinespace[2pt]
\textbf{DPE} & Image-only anchor & 72.4$\pm$1.7 & 65.1$\pm$2.9 \\
& CAPRA interface & 80.0$\pm$1.2 & 76.8$\pm$1.8 \\
& \shortstack[l]{CAPRA +\\ calibrated priors} & \textbf{81.3$\pm$1.0} & \textbf{78.5$\pm$1.5} \\
\midrule
\multicolumn{4}{l}{\textbf{HAM10000}} \\
\textbf{JTT} & Image-only anchor & 54.8$\pm$2.1 & 27.6$\pm$3.7 \\
& CAPRA interface & 69.5$\pm$1.9 & 41.0$\pm$2.9 \\
& \shortstack[l]{CAPRA +\\ calibrated priors} & \textbf{70.2$\pm$1.6} & \textbf{54.2$\pm$2.4} \\
\addlinespace[2pt]
\textbf{DPE} & Image-only anchor & 70.8$\pm$1.8 & 39.1$\pm$3.1 \\
& CAPRA interface & 74.5$\pm$1.4 & 49.4$\pm$2.3 \\
& \shortstack[l]{CAPRA +\\ calibrated priors} & \textbf{76.2$\pm$1.2} & \textbf{52.6$\pm$2.0} \\
\bottomrule
\end{tabular}
\end{table}

The clearest positive anchor is BRSET DPE, where replacing the image-only anchor raises WGA from 65.1 $\pm$ 2.9 to 76.8 $\pm$ 1.8. A second anchor appears on HAM10000 JTT, where calibrated priors increase WGA from 41.0 $\pm$ 2.9 to 54.2 $\pm$ 2.4. Together, these two gains separate the two roles of the interface: the interface itself makes subgroup structure explicit, while calibrated priors make that structure stable enough to exploit on noisier, longer-tail slices. Interface quality and posterior quality are therefore complementary rather than interchangeable.

This division of labor matters especially under missing metadata because the learner never observes the subgroup variables it would normally optimize against directly. The interface resolves a structural ambiguity problem by making hidden slices legible, while calibrated priors resolve a confidence-allocation problem by making those slices stable enough to influence optimization. The contrast between BRSET DPE and HAM10000 JTT is consistent with exactly that picture: one setting benefits first from making subgroup structure explicit, while the other benefits most once that structure has also been stabilized.

\section{Discussion}
CAPRA is best understood as a calibrated proxy-axis interface for hidden subgroup analysis under missing metadata, rather than a latent partitioning method. It does not aim to recover a fixed “true” group structure, but to make subgroup structure visible, interpretable, and reusable when labels are unavailable. This is critical in medical imaging, where deployment risks often remain hidden in aggregate metrics once metadata disappear. By organizing image variation into calibrated proxy-axis posteriors, CAPRA exposes these failure modes and provides a practical interface for subgroup analysis and optimization. This positioning is consistent with interpretable machine-learning decision support, where intermediate evidence should remain inspectable and actionable rather than only predictive \cite{li2019application}.
Its downstream gains are inherently domain-dependent. CAPRA is most effective when the baseline has not already captured dominant subgroup structure; when it has, the remaining gains are limited \cite{zong2022medfair,ricci2022addressing,xu2024addressing}. CAPRA should therefore be viewed not as a universally stronger learner, but as a mechanism for restoring subgroup visibility under missing metadata, with impact determined by how much additional structure it reveals beyond the baseline.

\textbf{Limitations.} CAPRA's proxy axes are not ground-truth metadata and should not be interpreted as exact recovery of sensitive or acquisition attributes. Sensitive-attribute proxies are intended for auditing and robustness analysis rather than direct clinical decision making. External validation also remains limited: mBRSET provides the clearest deployment-shift evidence, but denser multi-site validation is still needed to establish stability across broader hospital environments and acquisition protocols. Finally, the dominant failure axis changes across modalities and tasks, so there is no universal subgroup taxonomy that transfers unchanged across medical-imaging settings.

\textbf{Evidence Boundaries.} The current evidence supports calibrated subgroup-interface quality, retention of information under external shift, and domain-dependent downstream reuse. Stronger claims, such as broad statements about hidden-subgroup robustness, should wait for denser external validation, cleaner metadata protocols, and more matched-audit evidence.

\section{Conclusion}
CAPRA introduces a calibrated proxy-axis framework for hidden subgroup analysis under missing metadata. It exposes subgroup structure that remains informative under external shift, aligns discovered partitions with explicit failure axes, and emphasizes axes that are both reliable and failure-relevant. The resulting interface improves interpretability and can be reused by downstream robust learners, although the gains are domain-dependent.

More broadly, CAPRA shows that hidden subgroup analysis need not rely on opaque latent slices. By organizing image variation into calibrated proxy-axis posteriors, it provides a practical interface for subgroup auditing, failure localization, and downstream adaptation without access to metadata. Future work should strengthen the subgroup objective while maintaining this calibrated and interpretable interface, and evaluate it more extensively across diverse settings.

\bibliographystyle{ACM-Reference-Format}
\bibliography{ref}

\end{document}